\documentclass[a4paper,twocolumn]{esapub} % European paper
\usepackage{natbib}
\usepackage{graphicx} %[pdftex]
\usepackage{amsmath}
\usepackage{amssymb}

\newcommand{\um}{\ensuremath{\mu}m}

\newcommand{\uK}{\ensuremath{\mu}K}

\title{\protect{Microtraps and Atom Chips: Toolboxes for Cold Atom Physics}}
\author{L.~Feenstra\thanks{E-mail: Feenstra@PhysI.uni-Heidelberg.de}, L.M.~Andersson
and J.~Schmiedmayer}
\affil{Physikalisches Institut, Universit\"at Heidelberg, Philosophenweg 12, 69120
Heidelberg, Germany}

\begin{document}
%\DeclareGraphicsExtensions{.pdf,.png,.jpg,.mps}
\maketitle
\begin{abstract}
Magnetic microtraps and Atom Chips are safe, small-scale, reliable
and flexible tools to prepare ultra-cold and degenerate atom
clouds as sources for various atom-optical experiments.
We present an overview of the possibilities of the devices and indicate how a microtrap  can be
used to prepare and launch a Bose-Einstein condensate for use in an atom clock or an
interferometer.
\end{abstract}
\section{Introduction}
Some of the key elements in the sensitivity of interferometers and atom clocks are the free evolution
time and the collisional shift of the atom beams and cloud.
Using cold and ultra-cold atoms can reduce the atomic velocity down to cm/s, to increase the free-evolution
time.
This means an equivalent wavelength of the atom packet of approx.~0.1~\um.
However, at temperatures of several \uK\ the beam spread can still be considered large, and scattering
can take place inside the beam.
It is therefore interesting to further cool the samples and maybe even bring them to
degeneracy; a Bose-Einstein condensate (BEC) expanded to 1~mm diameter has a transverse velocity spread
on the order of \um/s.
Furthermore, reducing the density by thus expanding the cloud from $10^{13}$~cm$^{-3}$ to $10^{8}$~cm$^{-3}$
reduces the collisional shift of the atomic transitions to below 0.5~mHz~\citep{Fertig2000,Sortais2000}.

The outline of the paper is the following; we first present the building blocks of magnetic
microtraps and Atom Chips, next we briefly discuss the experimental state of the art.
Finally we suggest how the devices may be used for space applications.

A comprehensive review on trapping and manipulation of atoms with magnetic micro-traps has been
written by~\citet{fol02}.
Additional information on BEC in atom chips is given by~\citet{Rei02-469}.

\section{Wire traps and Atom Chips}

In order to minimise the footprint, weight and power consumption of magnetic traps for space
missions other routes than the traditional coil-based setups should be sought.
Surface mounted wire traps and Atom Chips have proven to be able to capture large numbers of atoms, and to
enable traps in various geometries.
By miniaturisation using microfabrication the traps can be made very steep and tight,
while maintaining mechanical stability, and robustness~\citep{fol02}.
These properties also ensure the reproducibility of the magnetic field parameters.
Combined with the inherent low inductance and fast reaction of small structures that need little
power the devices are ideal for use in sensitive environments.

\begin{figure}[ht]
\centering
\includegraphics[width=8cm,keepaspectratio]{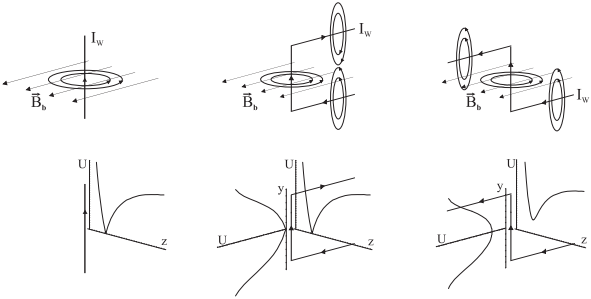}
\caption{Basic wire traps; the Side Guide, U-, and Z-trap. Top row: wire geometry and bias field. Bottom row:
resulting potential shape.}
\label{BasicWiretraps}
\end{figure}

The most basic micro-trap consists of a straight, current carrying wire in a homogeneous magnetic bias field
oriented perpendicular to it~\citep{Frisch33,fol02}, see Fig.~\ref{BasicWiretraps}.
The two superposed fields subtract to a 2D quadrupole field minimum along the wire, the Side Guide, in which atoms
can be held. The trap depth is given by the homogeneous
bias field, the field gradient is inversely proportional to the wire current.
An additional bias field oriented along the wire, effectively rotating the bias field,
yields a Ioffe-Pritchard trap with a non-zero trap minimum, which is robust against Majorana spin-flip losses.
Such a geometry lends itself to miniaturization of the wire size using micro-fabrication
techniques.
A typical example is the Atom Chip~\citep{Fol00-4749}, where
the 1~-~200~$\mu$m wide wires are micro-fabricated in a 2~-~5~$\mu$m thick gold layer on
a silicon substrate.
The present generations of Atom Chips used in our experiments, fabricated using a lithographic
technique~\footnote{The Atom Chips used in our experiments are fabricated in the Weizmann Centre of Science, Rehovot,
Israel}, are capable of carrying current densities of over
$10^7$~A/cm$^2$.

By bending the ends of the wire one creates slightly more complex
structures~\citep{Den99-291,Rei99-3398,Haa01-043405} that provide
three dimensional confinement, see Fig.~\ref{BasicWiretraps}.\\
The U-trap: by bending a wire in a U-shape, the two fields from the bent ends of the
wire (arms) close the initial two dimensional confinement (guide) along the central wire
segment (base).
The result is a three dimensional quadrupole field, with a trap minimum $B_{0}=0$.\\
The Z-trap: if the arms are pointing in opposite directions (Z-shaped wire) a field
component parallel to the base remains, resulting in a simple Ioffe-Pritchard type trap with $B_{0}>0$.

A common property of wire traps is the scaling of the strong linear confinement in the
transverse direction, inversely proportional to the wire current $I_W$ and the distance
$z_{0}$ between the wire centre and the trap minimum:
\begin{equation}
z_{0} \propto\frac{I_W}{B_{x}}, \hspace{0.7cm} B'_{\bot}=\frac{\partial B}{\partial x,z}
\propto\frac{{B}_{x}^2}{I_W}.  \label{randbo}
\end{equation}
Here $B_{x}$ is the homogeneous bias field perpendicular to the wire. In the case of a
Z-trap, the non-zero field minimum turns the bottom of the
quadrupole trap into a harmonic potential. The transverse angular trap frequencies
$\omega_{x,z}$ in this potential scale as:
\begin{equation}
\omega_{x,z} \propto \frac{B'_{\bot}}{\sqrt{B_0}}. \label{trapw}
\end{equation}

The linear field gradient of wire traps, given by Eq.~(\ref{randbo}),
enables a large flexibility to first trap a large number of atoms and second to
efficiently compress the trap to small volumes and high trap frequencies for effective
evaporative cooling.

\begin{figure}[h]
\centering
\includegraphics[width=7cm,keepaspectratio]{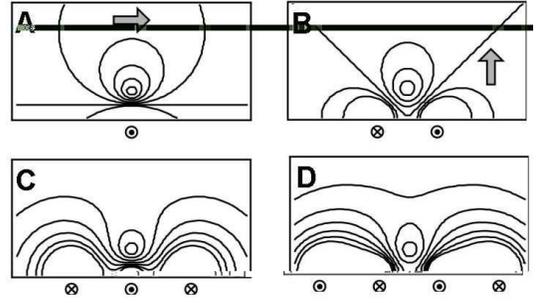}
\caption{Wire traps; variations.
A: Side guide with homogeneous field;
B: Two-wire guide with homogenous vertical bias field;
C: Side guide, bias field from additional wires;
D: Two-wire guide with bias field from additional wires.
}
\label{WireCombis}
\end{figure}

The bias fields need not be made by coils, they can
also originate from permanent magnets surrounding the setup.
The control of the current through the trapping wire maintains the flexibility of the device.
Another method is to supply the fields by currents in additional wires in parallel to the trapping wire,
 see Fig.~\ref{WireCombis}, or by the arms of the wire, as in the case of
 the longitudinal field in the Z-trap.

Since the trapping parameters of a Side Guide depend on the angle between the wire and the bias
field, bending the wire means to alter the trap.
This effect is absent in the geometry of two parallel wires carrying opposite currents in a vertical bias
field (Fig.~\ref{WireCombis}-B and \ref{WireCombis}-D), which is independent of the angle of the wires in the plane,
thus it allows atom guides in every direction on the chip.

\section{Experimental setup}\label{mountingdescription}

The starting point for cold atom experiments is the collection of a large number of atoms in an ultra-high vacuum
(UHV) chamber in a special magneto-optical trap, the Mirror-MOT.
A mirror-MOT is made with two sets counter-propagating laser beams.
One set is reflected off the chip at an angle of $45^\circ$, overlapping the axis of a magnetic quadrupole,
also tilted by $45^\circ$ with respect to the mirror surface.
The other set of beams runs in the plane of the chip~\citep{Lee96-1177,Pfau96,Rei99-3398,Fol00-4749}.
The magnetic quadrupole field for the MOT is supplied by external coils or by the field of the
U-trap (U-MOT).
Both setups can be operated with the same laser beam configuration.

For experiments with rubidium the laser light is provided by diode lasers, frequency locked to a spectroscopy signal.
The light is switched and tuned to the desired frequencies by acousto-optical modulators (AOMs)
and mechanical shutters.
The beams can be led to the experiment using mirrors or by optical fibres.
The latter option has the benefit of increased stability and a better intensity
distribution, however it also increases intensity losses.
The present studies on optical setups for the interferometers aboard the LISA- and SMART-2 missions,
as well as for the regular MOTs for HYPER and the atom clock PHARAOH/ACES will point the way to an
optimum optical setup for spaceflights.

\begin{figure}[h]
\centering
\includegraphics[height=3.5cm,keepaspectratio]{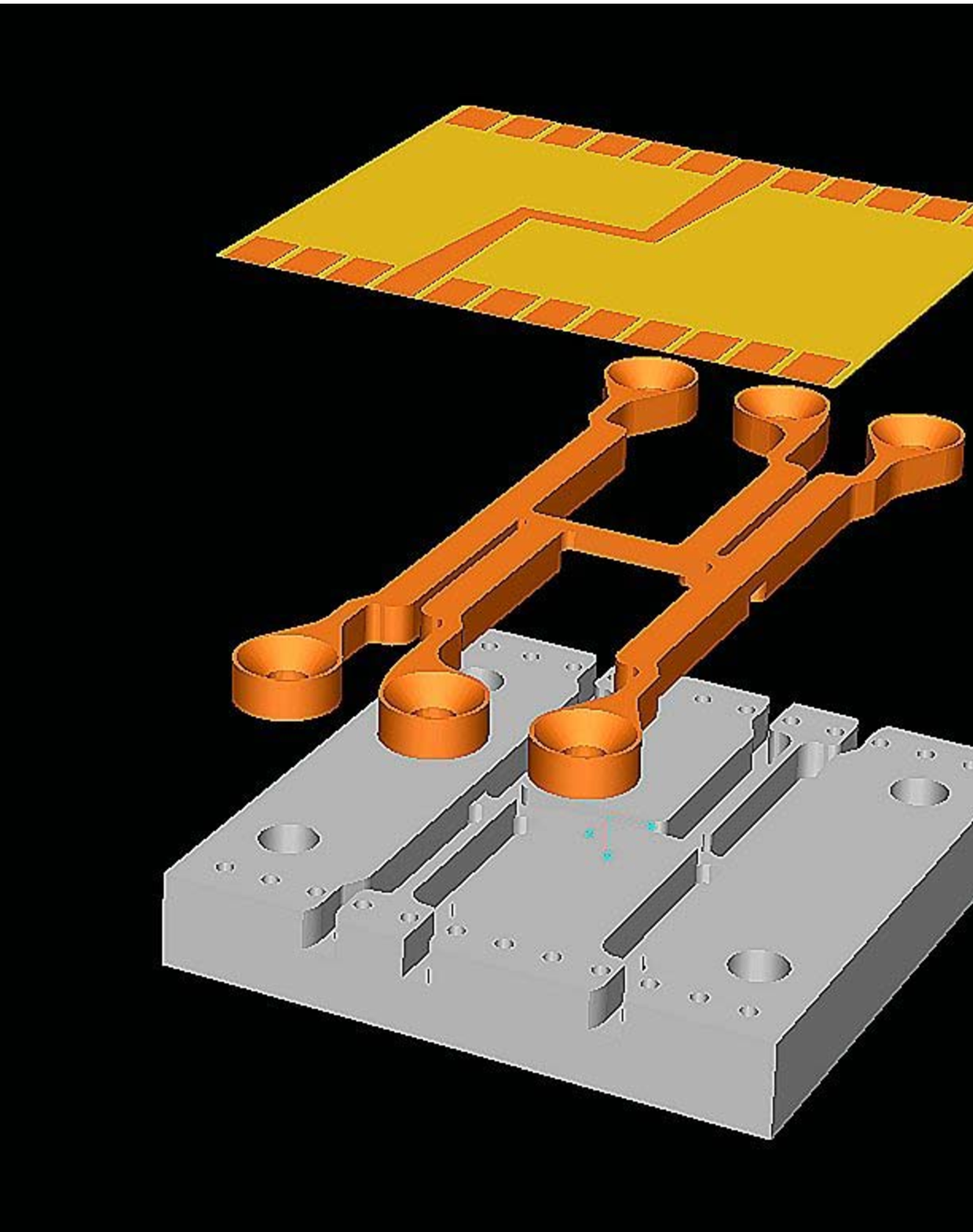} \quad
\includegraphics[height=3.5cm,keepaspectratio]{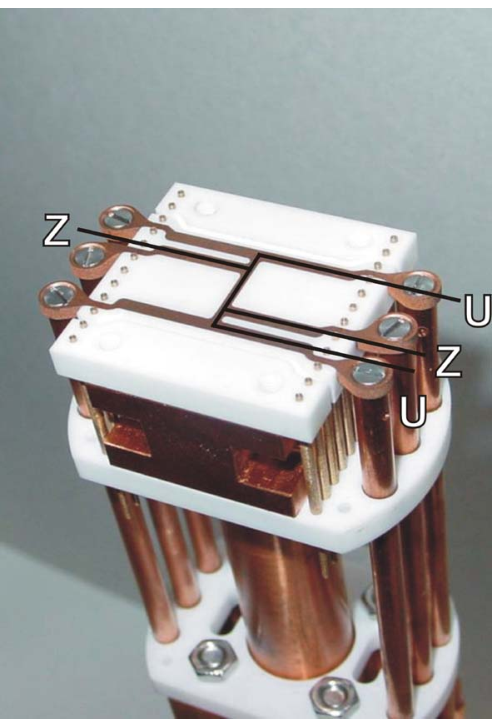}
\caption{The Atom Chip is mounted directly against a ceramic carrier, housing a copper
device for the initial, larger, traps. The structure is used for U- and Z-shaped currents
for quadrupole fields and Ioffe-Pritchard traps.}
\label{mounting}
\end{figure}

The heart of the experiment is a compact Atom Chip holder assembly, containing the Atom Chip
with the traps and guides for the desired experiments, and a copper structure embedded in
a UHV-compatible ceramic directly behind the chip, see Fig.~\ref{mounting}.
The copper part is designed to allow strong U- and Z-currents for large quadrupole and Ioffe-Pritchard traps.
The $1\times1$~mm$^2$ cross section withstands the typically used currents of 50~A for
over a minute without significant heating, thus preserving the UHV~\citep{Schneider2002,Kasper2002}.

The contacts to the Atom Chip are wire-bonded to contact-pins, with multiple bonds per
connection to reduce the contact resistance.
The maximum current through the chip-wires is limited by the dissipation of the heat produced in the chip,
not by the wire-bonds.
The copper devices are bolted to the leads for optimum conductivity.
For space flights the copper can easily be replaced by a superconducting
material to reduce the necessary power even further.

\begin{figure}[h]
\centering
\includegraphics[height=3cm,keepaspectratio]{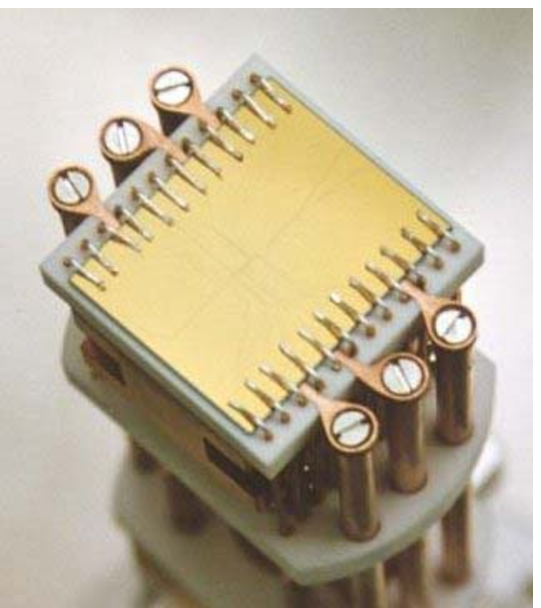} \quad
\includegraphics[height=3cm,keepaspectratio]{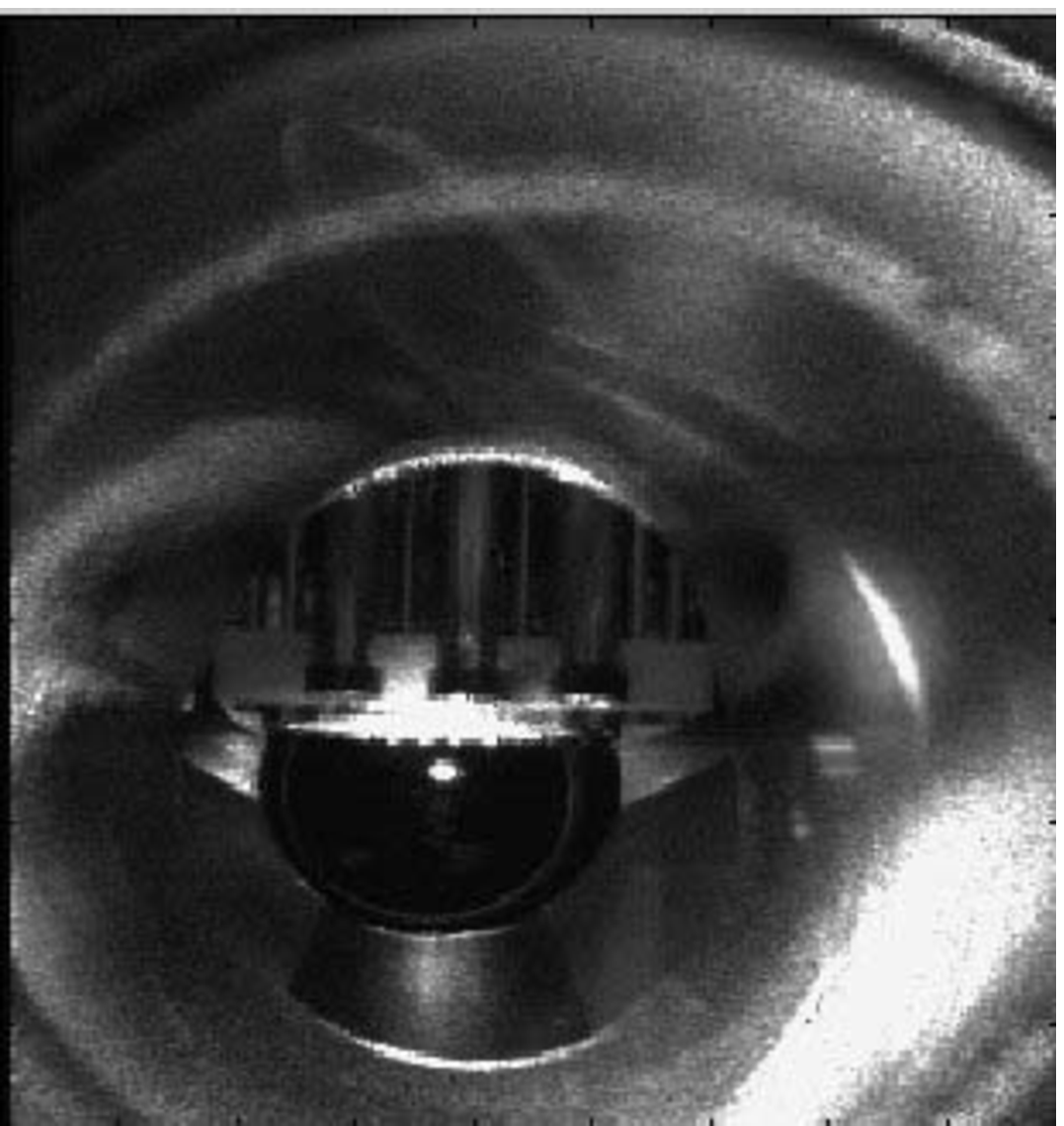}
\caption{Left: The mounted Atom Chip. Right: Rb Mirror-MOT in the UHV chamber, hanging under the chip.}
\label{MountedChipAndMot}
\end{figure}

An additional optimised U-structure underneath the copper H-structure can be used to generate the
initial quadrupole field for the mirror-MOT~\citep{BeckerDiplomarbeit}.
Since this removes the need for a separate set of quadrupole coils, it reduces the
complexity, weight and power consumption of the setup.
It also allows a much better optical access to the atoms for further manipulation and study.

The experiments can be performed in a single vacuum chamber~\citep{Ott01-230401,Haensel2001}.
In that case a switchable atom source and sufficient pumping power must be used to enable loading the
MOT from the background gas and to restore the vacuum after the MOT has been fully loaded.
Another option is to use a separate chamber to produce a beam of cold atoms to load the
MOT in the UHV-chamber~\citep{Schneider2002,Kasper2002}.
This has the advantage of enabling a much better vacuum at the location of the chip
and the experiments, which considerably extends the lifetime of the trapped clouds,
however it comes with the cost of a significantly increased complexity, weight and power consumption.

\section{Loading the Atom Chip}
A typical experimental cycle proceeds as follows:

1. Cooling and trapping of atoms in a Mirror-MOT\\
2. Further cooling via optical molasses\\
3. Optical pumping to the desired hyperfine state\\
4. Magnetic trapping of the atoms\\
5. Further cooling; trap compression and RF-cooling\\
6. Bose-Einstein condensation (if desired)\\
7. Transfer of the cloud to the chip\\
8. Manipulation of the atoms onboard the chip

Ad 1: In the final stages of the U-MOT-phase, the field
configuration is changed to compress the cloud and bring it closer to the chip, see Eq.~(\ref{randbo}).
This combines a large capture volume for the initial loading with a dense cloud for
optimum transfer to the magnetic traps.

Ad 4 -- 7: The initial magnetic trap may be directly from the chip, however
these traps are usually too shallow and too small to capture sufficient atoms to allow
evaporative cooling to a large BEC.
Therefore we use the copper structure behind the chip, which can be operated at much
higher currents and thus it can generate deep traps far away from the chip.
The trap can also easily be compressed which leads to higher collision rates inside the trap and thus to a faster
thermalisation rate of the sample, which is needed for efficient evaporative cooling~\citep{Davis95model,Luiten1996}.
Moving the copper trap closer to the chip enables ``mode matched'' transfer to the chip traps.

Ad~8: Once trapped onboard the chip the atom cloud can be transported
to other locations on the chip for use as a source for various experiments
either on the chip or in other devices.

\section{Use of microtraps and atom chips}
As shown by~\citet{fol02} and~\citet{Rei02-469} a large variety of traps and guides for cold atoms and Bose-Einstein
condensates has been realised.
The necessary properties of traps that should enable to reach condensation, such as the capability of
trapping large numbers of atoms, may not be possible in some experimental setups, such
as those aimed at building quantum dots.
Therefore, it is interesting to separate the parts with contradicting experimental demands into different parts of
the setup.
However, this also raises the need for transport of the atoms between the sites, as
discussed for the MOT above.
Used techniques range from mechanically moving the trapping coils~\citep{Jila} to transport in optical
traps~\citep{lea02-040401} and to both macroscopic~\citep{Greiner2001} and microscopic magnetic conveyor
belts~\citep{Haensel2001Motor}.

Our option is a combined multiple-layer structure of the type described above (Fig.~\ref{mounting}).
For instance, using a wire current of $I_W = 50$~A in the copper Z-trap and properly
tuned bias fields, thermal atoms can be trapped up to 3~mm from the wire center.
Such large distances easily allow the micro-fabricated structures of Atom Chips to be
placed between the wire and the trap.

Next we discuss two main uses of our traps: the creation of a large cold cloud or a sizable
Bose-Einstein condensate and the manipulation and preparation of cold clouds for beam splitting and
interferometry.

\subsection{Bose-Einstein condensation}
\begin{figure}[h]
\centering
\includegraphics[width=8cm,keepaspectratio]{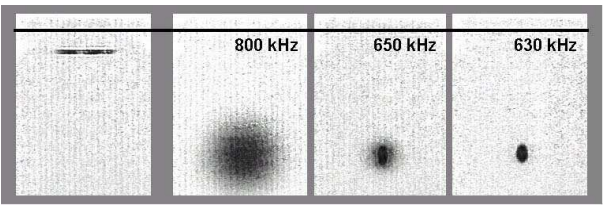}
\caption{Creating a BEC and loading it in the Atom Chip: \emph{Far left}: Atoms in the final trap,
\emph{Middle left to right}: Growth of the BEC upon lowering the temperature (T $\propto$ RF-Frequency)
(after 16~ms expansion). Each image displays $\sim\!1.8\times2.3$~mm$^2$.
The chip surface is indicated with the black line
}
\label{BEC}
\end{figure}

To achieve Bose-Einstein condensation in $^{87}$Rb, we start with the magnetic Z-trap of the copper structure.
By varying the bias fields from 25~G to 60~G, the trap is compressed in 19~seconds to field gradients in excess of
400~G/cm at a trap distance of 1.5~mm from the wire center, reaching trap frequencies of
$\omega_{\perp} > 600$~Hz, and reducing the trap minimum from 5~G to $\sim\!1$~G.
This linear compression ramp is combined with evaporative cooling by applying
a linear RF-frequency ramp from 19~MHz to ca.~0.6~MHz.
The resulting Bose-Einstein condensate contains approx.~$3\times10^5$~atoms in the
$\left|{\rm F}=2,\mathrm{m_F}=2\right>$ state~\citep{Schneider2002,Kasper2002}, located approx.~250~\um\
from the chip surface.
Fig.~\ref{BEC} shows the atoms in the final trap, and the growth of the condensate
in absorption images after 16~ms of expansion.

The condensate can be transferred to a 2~mm $\times$ 200~\um\ (base length $\times$ wire width) Z-trap
 on the chip in typically 100~ms.
Since the condensation procedure does not use any chip trap complete freedom remains
in the design of a specific wire layout on the chip.

Using wire cross sections as low as $1\times1$~\um$^2$ we have achieved trap frequencies of over 1~MHz.
Such extremely tight traps enable very fast thermalisation, which may be used for rapid
condensation on site aboard the chips.
For instance: condensation on an Atom Chip in less than a second has been shown
by~\citet{Rei02-469}.

In the present experiments the chips have not been cooled.
Theory predicts however, that decoherence can result from thermal induced current noise
in the trapping wires, which effect becomes more important upon closer approach to the wires~\citep{Henkel2002}.
It is therefore interesting to investigate the influence on the coherence properties of cold samples
when parts of the setup or the setup as a whole are cooled.

\subsection{Beam splitting and interferometry}
Combining several trapping and guiding geometries, beam splitters and interferometers can be built
on board an Atom Chip.
One can divide between spatial and temporal interferometers.
Traps can be split magnetically~\citep{fol02,Rei02-469}, electrically~\citep{fol02,KleinDiplomarbeit} and
optically~\citep{Dav95-3969,Greiner2002Mott}.

As an example we present the case of two wires with co-propagating currents in a transverse
bias field, which may be used for temporal interferometry~\citep{Den99-291,Hinds2001}.
At zero bias field, there exists a magnetic quadrupole minimum in the plane of the wires, which in
itself may be used to guide atoms~\citep{Thywissen99EPJD,Sauer2001}.
When a transverse bias field is added, this in-plane minimum is shifted upwards while far
away from the wires another minimum forms according to Eq.~(\ref{randbo}) with $I_W$ the
total current of both wires.
For increasing bias field strengths the minima approach each other along the
symmetry plane until they merge in a single trap at a critical bias field strength.
For even higher bias fields the finite-size effect of the trapping currents running in two separate wires instead
of in a single wide wire prevails and the trap splits in two minima that move towards the wires.

\begin{figure}[h]
\centering
\includegraphics[width=6cm,keepaspectratio]{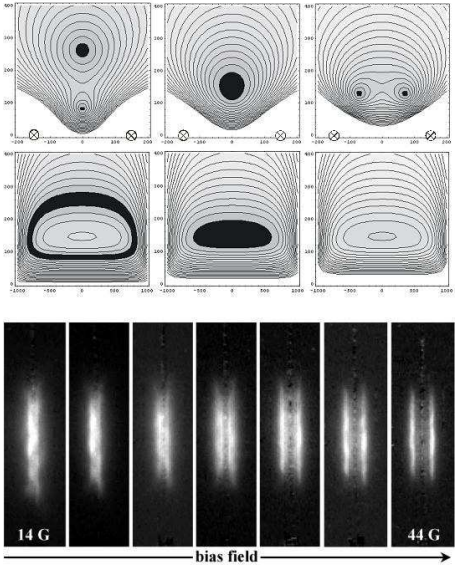}
\caption{Transverse splitting of an atom cloud. Wire currents: 2~A, for other details: see text.}
\label{transsplit}
\end{figure}

Fig.~\ref{transsplit} shows the equipotential lines in the symmetry plane between the wires and the experimental
results of this procedure for two U-shaped wires, arranged as $\sqsupset\,\sqsubset$, with
the currents co-propagating along the bases~\citep{fol02,WildermuthDiplomarbeit}.
The atoms are first trapped in the outer minimum, upon increasing the bias field the trap is fully split.
It is interesting to note that with the shown wire layout the horizontal splitting is
complete as intended, however the vertical splitting results in a closed-loop atom guide.
Since the open area can be controlled via the bias fields, this effect might be useful for experiments.

\begin{figure}[h]
\centering
\includegraphics[width=4.5cm,keepaspectratio]{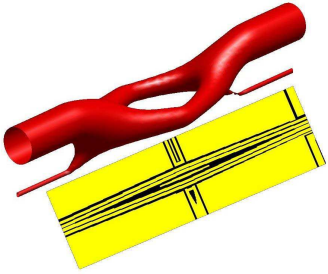}
\caption{Wire layout of a spatial atom interferometer (with phase shifters halfway)
and the resulting potential landscape.}
\label{IFM}
\end{figure}

For spatial interferometry two X- or Y-shaped beam splitters can be put back to back.
Beam splitting with magnetic Y-geometries has been shown and discussed in~\citep{Cassettari2000,Cas00-5483}.
The coherence characteristics of a Y-beam splitter and of an atom interferometer built from two
Y-splitters (Fig.~\ref{IFM}) are discussed in detail by~\citet{Andersson2002}.
In addition, a single X-beam splitter can be used as a Michelson interferometer~\citep{Dumke2002Lines}.

\section{Possible applications for use in space}
The use of on chip interferometer designs has the advantage of
well defined interferometer arms.
However, to reach the extreme sensitivity on the order of $10^{-18}$, aimed for in precision interferometry experiments,
ultra-stable control fields are needed.
To use an Atom Chip interferometer for such precise measurements many questions still have to be addressed.
This is an active research field, but at present there are no studies on
systematic effects in guided matter wave interferometers.
On the other hand, an Atom Chip setup is very well suited to serve as a source of cold atom clouds with precisely
determined characteristics for use in free space interferometers.

By lowering the launch velocity of the cold atom cloud used in an interferometer,
the Bragg diffraction in the beam splitter leads to a larger deflection angle.
For a wavelength of 780~nm the recoil velocity for $^{87}$Rb atoms is $\sim\!6$~mm/s.
If the atoms have an equal forward velocity the deflection angle
is $45^\circ$ and a very large area can be enclosed by the interferometer.
In order to benefit from this low velocity the momentum uncertainty in the cloud
has also to be low.

With the magnetic Z-trap the BEC is prepared close to the chip surface.
By smoothly accelerating the condensate while ramping down the trap frequency
a slow free flying condensate with a long de~Broglie wave length can be produced.
The procedure can easily be adapted to deliver any cold atom cloud with specifically tailored
properties for optimum use in atom clocks and interferometers.

\begin{figure}[!b]
\includegraphics[width=8cm]{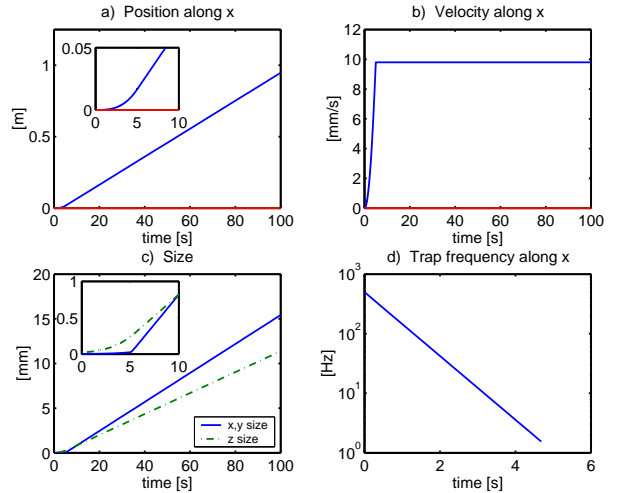}
\caption{Simulation of the evolution of a condensate during and after the acceleration and
expansion in an atom chip trap. a) The center of mass position. b) The center of mass velocity.
c) Condensate size evolution. d) Transverse trap frequency ramp, maintaining the initial aspect ratio of 10.
The long axis is in the $z$ direction and the acceleration out of the chip is in the $x$ direction. \label{fig:launch}}
\end{figure}

Fig.~\ref{fig:launch} shows the results of a simulation of the release of a BEC with a final
velocity of about 10~mm/s.
The theory for the BEC dynamics in harmonic traps, in which the center of mass motion decouples
fully from the shape dynamics, has been discussed by~\citet{Kagan97} and~\citet{Japha02}.
A further assumption is that the condensate is in the Thomas-Fermi regime.
The initial condensation trap has frequencies of $500~\rm{Hz}\times500~\rm{Hz}\times50~\rm{Hz}$
in the $x,y,z$ coordinates.
The initial BEC is assumed to consist of $10^5$ atoms. The acceleration is applied for 5 seconds
with a constant rate of change of the acceleration to ensure a smooth motion (see also \citet{Gustavson2002}).
After this time the trap centre is about 15~mm away from the starting point, which is easily achieved with
the presented wire traps in micro-gravity conditions.
At the same time the trap frequencies are ramped down exponentially to ensure
a more or less adiabatic evolution of the condensate. The result is a condensate
travelling at a speed of $\sim\!10$~mm/s, thus moving $\sim\!1$~m in 100~s.
Due to the almost adiabatic expansion, the expansion velocity of the condensate is lowered and
after a 100~s flight the condensate is approximately spherical with
a diameter of only approx.~15~mm. Thus the scheme provides an excellent source for
cold atom clouds for use in an atom interferometer.

\section*{Acknowledgements}
This work was supported by the EU project ACQUIRE, Austrian Science Fund (S065-05, SFB F15-07)
and the Deutsche Forschungsgemeinschaft
(grant Schm1599/2-1). L.F. acknowledges the support of the Alexander von Humboldt Foundation.
\bibliography{hyperarxiv}
\end{document}